\documentclass[journal]{IEEEtran}

\usepackage{cite}      
\usepackage{siunitx}
\usepackage{array}
\usepackage{amssymb,amsmath}
\usepackage{tabularx}
\usepackage{float}
\usepackage{subcaption}
\usepackage{graphicx}
\usepackage{grffile}

\makeatother

\author{Gopalakrishnan Sundararajan,~\IEEEmembership{Student~Member,~IEEE}~and~Chris Winstead,~\IEEEmembership{Senior~Member,~IEEE}%
	\thanks{This work was supported by the US National Science Foundation under award ECCS-0954747, and by the Franco-American Fulbright Commission for the Exchange of Scholars. G.\  Sundararajan was supported by Sant Graduate Innovation Fellowship at Utah State University.} %
	\thanks{G.\ Sundararajan and C.\ Winstead are with the Department of Electrical and Computer Engineering, Utah State University, Logan, UT 84322-4120. Email {\em gopal.sundar@aggiemail.usu.edu} and {\em chris.winstead@usu.edu}.} %
}

\begin{document}
	\title{ASIC Design of a Noisy Gradient Descent Bit Flip Decoder for 10GBASE-T Ethernet Standard}
	
	\maketitle

 \begin{abstract}	
In this paper, the NGDBF algorithm is implemented on a code that is deployed in the IEEE 
802.3an Ethernet standard. The design employs a fully parallel architecture and operates in two-phases: start-up phase and decoding phase. The two phase operation keeps the high latency operations off-line, thereby reducing the decoding latency during the decoding phase. The design is bench-marked with other state-of-the-art designs on the same code that employ different algorithms and architectures. The results indicate that the NGDBF decoder has a better area efficiency and a better energy efficiency compared to other state-of-art decoders. When the design is operated in medium to high signal to noise ratios, the design is able to provide greater than the required minimum throughput of \SI{10}{Gbps}. The design consumes \SI{0.81}{ \milli\meter^2} of area and has an energy efficiency of \SI{1.7}{\pico\joule/bit}, which are the lowest in the reported literature. The design also provides better error performance compared to other simplified decoder implementations and consumes lesser wire-length compared to a recently proposed design.
 \end{abstract}
 
 \section{Introduction}
 \label{Intro}
 
Low Density Parity Check (LDPC) codes were introduced by Gallager in 1963 \cite{Gallager_1963}. Since their  reintroduction by MacKay and Neal, LDPC codes have gained a lot of attention in the information theory community \cite{MacKay_1997a}. Due to very high decoding performance, many classes of LDPC codes have been constructed and a wide spectrum of decoding algorithms have been proposed to decode LDPC codes. 
 
The performance of LDPC codes are determined by the decoding algorithm used to decode the corrupted received bits from the channel. Decoding algorithms are iterative in nature and the decoding is done by iteratively exchanging messages between the two sets of nodes represented in a Tanner graph. Gallager's decoding algorithms can be classified into two main categories: hard-decision Bit-Flip Algorithms (BFA), and the soft-decision Sum-Product Algorithm (SPA).  Among these algorithms, SPA typically shows the best performance but suffers from high implementation cost because of high computational complexity. A variety of approximate SPA-based algorithms have been developed, including the Min-Sum (MS), Offset MS (OMS) and Normalized MS (NMS). These approximate algorithms are much less complex than the original SPA, but still require very sophisticated implementations. BFA, by contrast, has very low complexity and is easier to implement in hardware but suffers from low error correcting performance.
 
In the last decade, a new class of low-complexity algorithms have been proposed which bridge the performance/complexity gap between SPA and BFA \cite{Kou_2001,Zhang_2004}. These new set of algorithms are known collectively as Weighted Bit Flipping (WBF) algorithms. WBF algorithms use soft channel information to manipulate hard decisions during the decoding process. These algorithms are similar in complexity to BFA, while it employs reliability information similar to SPA algorithms. Hence, they offer a good trade-off between performance and complexity. Many variants and modifications to the WBF algorithm have been reported to date \cite{wu2006fast,Jiang_2005}. 
 
The WBF algorithms employ an inversion function local to each symbol node that determines whether the associated symbol is flipped. The inversion functions account for both the local soft channel information and the adjacent parity-check information, which is updated in each iteration. WBF algorithms may adopt sequential flipping mode, in which a single symbol is flipped in each iteration by identifying the symbol with the lowest inversion function metric. They may alternatively adopt parallel flipping, where multiple symbols are simultaneously flipped by applying a threshold operation. Sequential flipping  tends to offer better performance, whereas parallel flipping offers lower complexity and greater speed. Parallel flipping algorithms are often referred to as ``multi-bit flipping''.
 iterations \cite{Mobini_2007}. The DD-BMP algorithm was shown to be an effective low complexity alternative to the SPA algorithm.
 
Wadayama et al.\ formulated a novel inversion function based on the gradient descent formulation \cite{Wadayama_2010_TCOMM}. The algorithm, known as Gradient Descent Bit Flipping (GDBF), outperforms the original WBF algorithm. GDBF also converges faster and employs low latency arithmetic operations compared to the original WBF algorithm. Many variants of GDBF have been proposed to either improve the performance or reduce the complexity \cite{Ismail_2010,Phromsa_2012}. A major drawback of the GDBF algorithm is that it still suffers from lower performance compared to the SPA algorithm and its variants. This is because the GDBF algorithm tends to become trapped in local maxima.
To counter the undesired effects of these local maxima, Wadayama et al.\ devised a GDBF algorithm with escape process that provides an escape from the local maxima. This algorithm, which we refer to as Hybrid-GDBF (H-GDBF), performs repeated mode switching between parallel and sequential flipping modes by evaluating a global objective function. The objective function calculation is a high latency global operation that needs to be done over an entire code length, thereby rendering the high speed realization of H-GDBF difficult. 
 
As an alternative to H-GDBF, the authors recently proposed adding random perturbation during the decoding process to aid in escaping from local maxima \cite{Sundararajan_2014_TCOMM}. The algorithm termed Noisy Gradient descent Bit flipping (NGDBF), adds an independent Gaussian distributed random noise perturbation to the inversion function in each symbol and at every iteration. The noise perturbation results in significantly improved performance. In this paper, we describe an NGDBF implementation for the IEEE 802.3an standard that comes within \SI{0.2}{\decibel} of a benchmark OMS design from recent literature. Our design has the lowest area reported for this standard, as well as the lowest power consumption, and consumes the least energy per bit when operated at $E_b/N_0$ of at least \SI{5.5}{\decibel}. Compared to previously reported OMS decoders, the NGDBF decoder consumes lower energy per bit by $3.47\times$ (when compared to a weak-performing split-row decoder) to $33.9\times$ (when compared to a high-performance OMS decoder).

The rest of this paper is organized as follows: Section \ref{rel} provides a review of related work, and Section \ref{review} describes the notation and operation of the NGDBF algorithm. The specific details related to the NGDBF algorithm on the 10GBASE-T code are described in section \ref{sec:Implement}. Section \ref{sec:Arch_IEEE_802.3an} describes the hardware architecture of our design. Section \ref{sec:Results} provides implementation results, performance analysis and benchmark comparisons. Conclusions are described in Section \ref{Conclude}.

\section{Related Work}
\label{rel}
 
The recent work on $10$GBASE-T Ethernet LDPC decoder designs could be characterized by two key aspects: The first one is the algorithm implemented and the second one is the architecture. In this regard, five main designs are reviewed in this section. The first one is the offset MS decoder that was fabricated on a \SI{65}{\nano\metre} CMOS process \cite{Zhang_2010a}. This decoder employs a partially parallel and a pipe-lined architecture that provides a moderate throughput. This design is very complex, incurs a large area and has a high energy consumption. However, because of the superiority of the offset MS algorithm, it provides a very good error performance.

The next design that was proposed is the fully parallel split-row MS algorithm \cite{Mohsenin_2010a}. This design implements a low complexity version of the Normalized MS algorithm that significantly reduces the routing complexity. The key idea behind the split-row MS algorithm is to partition the original parity check matrix into many sub-matrices, thereby splitting a row processing operation, into multiple row processing operations. Check node computations for each sub-matrix are performed separately, using limited information from other columns. This reduces the routing congestion as it reduces the number of wires between the row and the column processors. However, when the original matrix is broken into 16 sub-matrices, there is a significant performance degradation of $0.35$ $\si{\deci\bel}$. This design consumes less area compared to the offset MS decoder and is highly energy efficient compared to the offset MS decoder.

Cevrero et al. proposed a layered implementation of the offset MS algorithm \cite{Cevrero_2010}. In this design, original parity check matrix of the $10$GBASE-T code is split into six layers in which each layer has $64$ rows and $2048$ columns. This enables the check node operation to be time multiplexed and the check node processor to be shared across layers. Since only $64$ check node processors are active at a time, only $64$ check node processors are needed to accomplish successful decoding. Another advantage of the layered implementation is that it provides faster convergence. This design was fabricated in a \SI{90}{\nano\metre} CMOS and achieved a throughput close to the required specification. This design consumes more than a watt of power and is inferior to Zhang's offset MS decoder in terms of energy efficiency.

As an alternative to the offset MS decoding, Tehrani et al. proposed a stochastic Majority-based Tracking Forecast Memory (MTFM) based $10$GBASE-T Ethernet LDPC decoder \cite{Tehrani_2010}. This design was done in a \SI{90}{\nano\metre} process. This design uses a fully parallel architecture and consumes a smaller area compared to both the offset MS and the split-row MS designs. This design has a better error performance than the split-row MS design. However, the stochastic MTFM decoder still has a significant complexity due to the requirement of a large number of random number generators to convert probabilities to streams of random numbers. All of the above mentioned designs, employ algorithms that are variants of the BP algorithm and are complex.

As an energy efficient alternative to the stochastic decoder, Cushon et al. recently proposed a low complexity design that employs a binary message passing algorithm and was implemented in a \SI{65}{\nano\metre} CMOS process\cite{Cushon_2014}. The algorithm termed Improved Differential Binary (IDB) consists of simple check and symbol node operations. IDB is a variant of the Modified Differential Decoding-Binary Message Passing (MDD-BMP) algorithm in which binary message passing is employed and is much simpler compared to the MS and the stochastic decoding algorithms \cite{Mobini_TCOMM_2009}. MDD-BMP performs very poorly on the $10$GBASE-T code. So, the authors proposed two modifications to the MDD-BMP algorithm to improve its performance. They are degeneration and relaunching. Degeneration is method in which the symbol node function is modified to enable the MDD-BMP escape the effects of weak absorbing sets. Relaunching is a technique in which failed frames are decoded in successive attempts, with subtle changes to the initial state of the decoder. These changes are deterministic and are based on a look-up table. The IDB decoder employs a fully parallel architecture and renders a very high throughput. It also has a better error correcting performance than the split-row MS algorithm and is the most efficient in terms of area consumption and energy dissipation compared to all other previous $10$GBASE-T decoders.
 
\section{Multi-bit Noisy GDBF (M-NGDBF) Algorithm}
\label{review}
 
\subsection{Notation}
Let $H$ be a binary $m\times n$ parity check matrix, where $n > m \ge 1$. To $H$ is associated a binary linear code defined by $C \triangleq \left\{ c \in F_{2}^n : Hc = 0\right\}$, where $F_{2}$ denotes the binary Galois field. The set of bipolar codewords, $\hat{C}\subseteq \left\{-1,\,+1\right\}^n$, corresponding to $C$ is defined by $\hat{C} \triangleq  \left\{\left(1-2c_{1}\right), \left(1-2c_{2}\right), ..., \left(1-2c_{n}\right) : c \in C \right\}$. 
 
Symbols are transmitted over a binary input AWGN channel defined by the operation $y = \hat{c} + z$, where $\hat{c} \in \hat{C}$, $z$ is a vector of independent and identically distributed Gaussian random variables with zero mean and variance $N_0/2$, $N_0$ is the noise spectral density, and $y$ is the vector of samples obtained at the receiver. We define a decision vector $x\in\left\{-1,+1\right\}^n$. Let $x\left(t\right)$ be the hard decision vector at a specific iteration $t$, where $t$ is an integer in the range $\left[0,\, T\right]$ where $T$ is the maximum number of iterations permitted by the algorithm. The decision vector is initialized as the sign of received samples, i.e.\ $x_k\left(t=0\right) = {\rm sign}\left(y_k\right)$ for $k=1,\,\dots,\,n$.
 
The parity-check neighborhoods are defined as $N\left(i\right)\triangleq\left\{ j : h_{ij} = 1 \right\}$ for $i=1,\,\dots,\,m$, where $h_{ij}$ is the $ij$ element of the parity check matrix $H$. The symbol neighborhoods are defined similarly as $M\left(j\right) \triangleq \left\{ i : h_{ij} = 1 \right\}$ for $j=1,\,\dots,\,n$.  The code's parity check conditions can be expressed as bipolar syndrome components $s_i\left(t\right) \triangleq \prod_{j\in N(i)}x_{j}\left(t\right)$ for $i=1,\,\dots,m$. A parity check node is said to be {\em satisfied} when its corresponding syndrome component is $s_i = +1$.

\begin{figure}[!t]
	\begin{center}
		\includegraphics{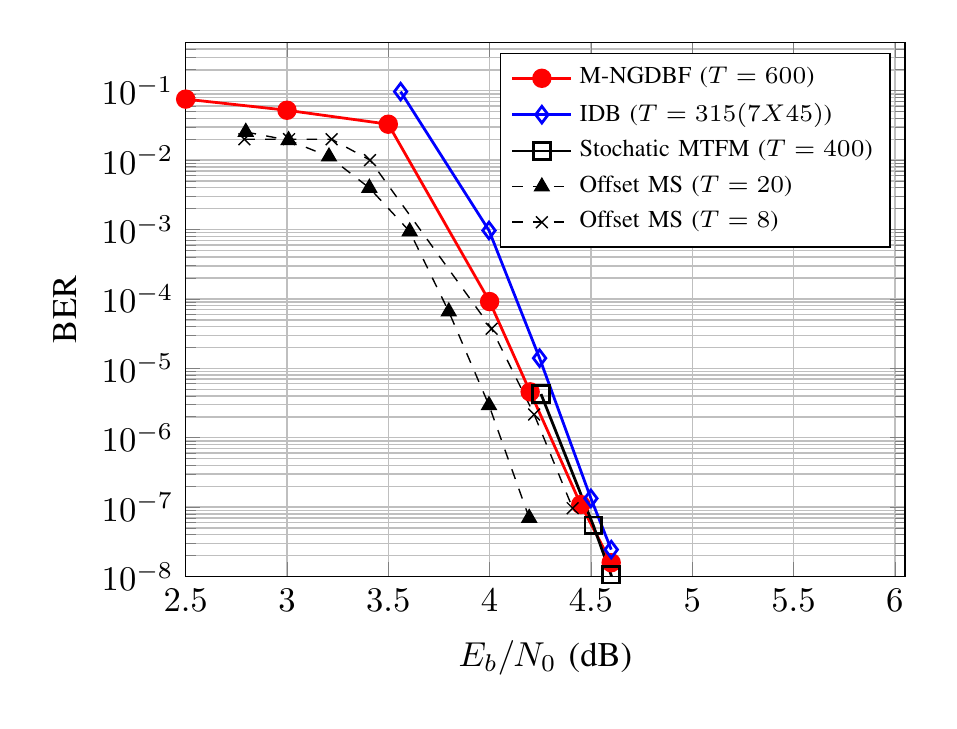}		
		\caption{ BER for NGDBF compared to a benchmark OMS decoder for the IEEE 802.3 standard LDPC code with maximum iterations limited to $T$.}
		\label{Fig:BER_IEEE802.3an}
	\end{center}
\end{figure}
  
\begin{figure}[!t]
   \begin{center}
	\includegraphics{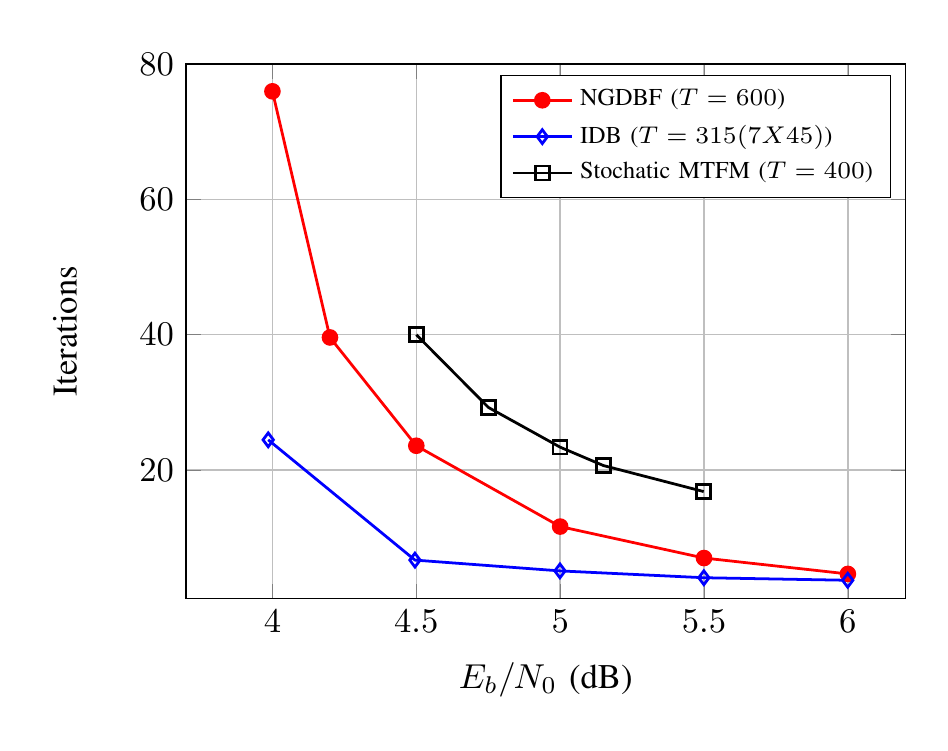}		
        \caption{ Average Number of Iterations for NGDBF algorithm for IEEE 802.3 standard LDPC code with maximum iterations limited to $T$.}
        \label{Fig:Avg_IEEE802.3an}
   \end{center}
\end{figure}

\subsection{Algorithm}
\label{sec:algorithm}

The GDBF algorithm proposed in \cite{Wadayama_2010_TCOMM} was derived by considering the maximum likelihood problem as an objective function for gradient descent optimization. In order to include information from the code's parity check equations, the syndrome components are introduced as a penalty term, resulting in the following objective function:
\begin{equation}
f\left(\mathbf{x}\right) = \sum_{k=1}^{n} x_ky_k + \sum_{i=1}^{m}s_i. \label{eq:objective_function}
\end{equation}

By taking the partial derivative with respect to a particular symbol $x_k$, the local inversion function corresponding to the GDBF algorithm is obtained as follows:
\begin{equation}
E_k = x_k\frac{\partial f\left(\mathbf{x}\right)}{\partial x_k} = x_ky_k + \sum_{i\in{\mathcal{M}}\left(k\right)} s_i. \label{eq:sym_func}
\end{equation}

\noindent In previous work, the authors modified the inversion function (\ref{eq:sym_func}) by adding a Gaussian distributed random noise sample as a perturbation term \cite{Sundararajan_2014_TCOMM}. The resulting Multi-bit NGDBF (M-NGDBF) algorithm can be summarized as follows:
 
 \begin{enumerate}[\setlabelwidth{Step 4}]
 	\item[\textbf{Step 1:}] Compute syndrome components $s_i = \prod_{j\in \mathcal{N}\left(i\right)}x_{j}$, for all $i \in \left\{1, 2,...., m\right\}$. If $s_{i}=+1$ for all $i$, output $x$ and stop.
 	\item[\textbf{Step 2:}] Compute inversion functions. For $k \in \{1,\,2,\,\dots,\,n\}$ compute\newline
 	$E_{k} = x_{k}y_{k} + w_{k}\sum_{i\in \mathcal{M}\left(k\right)} s_i + q_{k}$\newline
 	where $w_{k}$ is a syndrome weight parameter and $q_{k}$ is a Gaussian distributed random variable with zero mean and variance $\sigma^2 = \eta^2N_{0}/2$, where $0 < \eta \leq 1$.  All $q_k$ are independent and identically distributed.
 	\item[\textbf{Step 3:}] Bit-flip operations. Flip any bits for which $E_{k} < \theta$, where $\theta \in \mathbb{R}^{-}$ is the {\em inversion threshold.} 
 	\item[\textbf{Step 4:}] Repeat steps 1 to 3 till a valid codeword is detected or maximum number of iterations is reached.
 \end{enumerate}
 
 The parameters for this algorithm, namely the syndrome weight $w_k$, the noise scale $\eta$, and the threshold $\theta$ are determined empirically and are chosen to minimize the error rate (BER). For regular codes, including the 10GBASE-T code, a single weight parameter can be used for all symbol nodes, in which case the subscript $k$ is omitted.

\begin{figure*}[!t]
  \centering 
  \hspace*{-0.3cm}     
  \includegraphics[scale = 1.5,width=13cm,height=13cm]{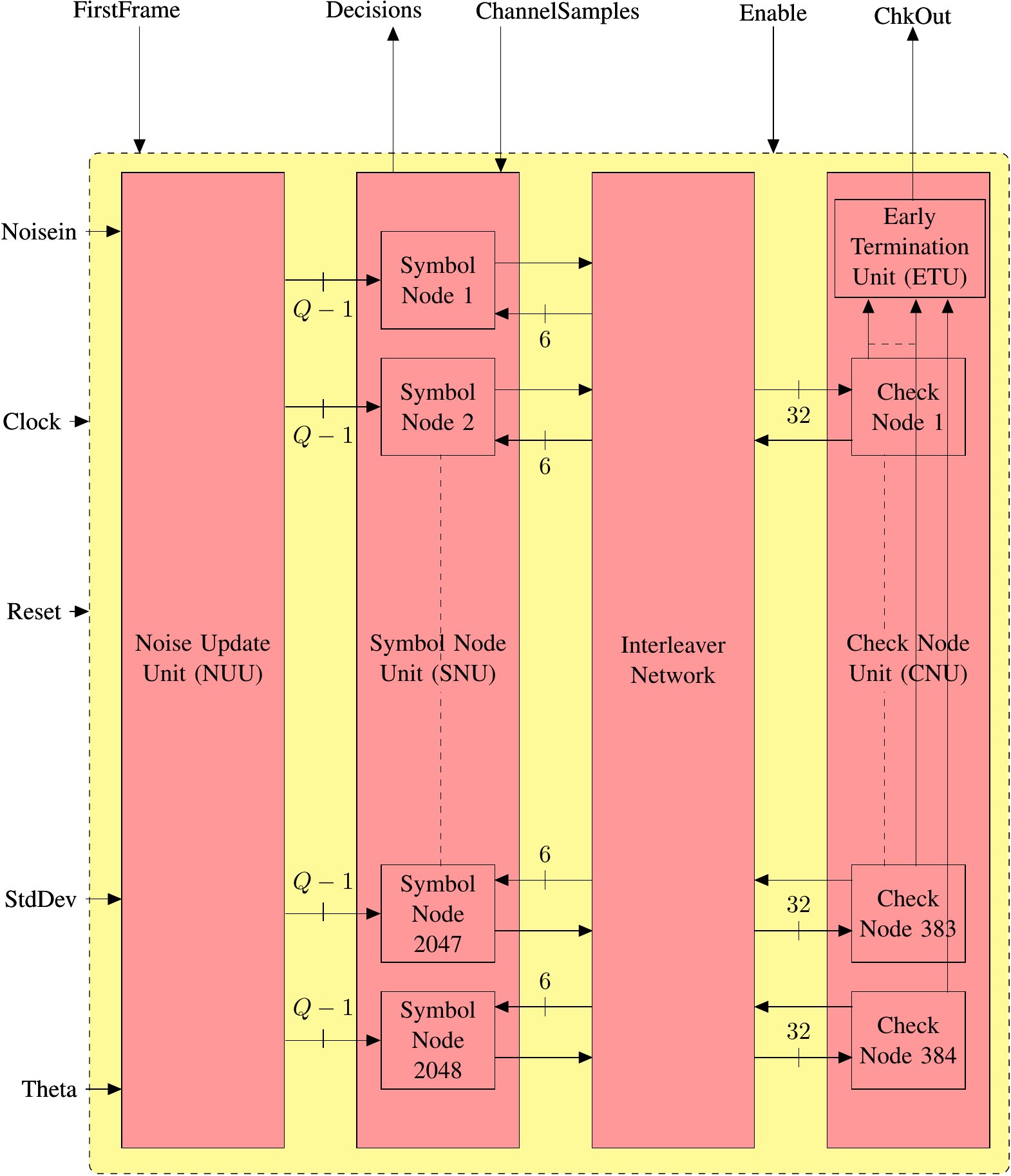}
  \caption{Top level decoder architecture showing all the blocks. Input, output and all the control signals are also clearly shown. $d_{v}(6)$ messages arrive at the symbol node from the interleaver and $d_{c}(32)$ messages arrive at a check node from the interleaver during a decoding iteration. Symbol node update requires $Q-1$ bits of noise from the $NUU$ every iteration.}
  \label{Fig:Top_Arch}
\end{figure*}
 
\section{Implementation}
\label{sec:Implement}

This section details some specific algorithmic parameters related to the NGDBF algorithm for IEEE 802.3an 10GBASE-T Ethernet Standard. The code deployed in this standard is a Reed-Solomon (RS) LDPC code and is also a common benchmark code for highly parallel implementations \cite{IEEE802.3_online}.  The code has a regular (6,32) degree distribution and the code rate is $0.841$. The parity check matrix corresponding to this code has $2048$ columns and $384$ rows. For this code, performance simulations show that the more complex NGDBF heuristics discussed in \cite{Sundararajan_2014_TCOMM} (namely threshold adaptation and sliding window smoothing) are not needed; the simple algorithm described in Section \ref{sec:algorithm} obtains an error performance that is comparable to other state of the art decoders reported on the IEEE 802.3an 10GBASE-T Standard. The chosen syndrome weight is $w=0.166$ $(1/6)$. The inversion threshold is chosen to be $\theta=-0.55$. The magnitude of channel samples was saturated at $2.95$. All the above mentioned parameters are estimated empirically from simulations to optimize the bit error rate (BER) performance. To further simplify the design, we employ {\em sample reuse} by cyclic-shifting the noise samples ($q_k$) used in Step 2. This reduces the requirement for random number generation, thereby leading to a very efficient design without impacting performance.

Fig. \ref{Fig:BER_IEEE802.3an} shows the performance of the NGDBF algorithm in comparison with other algorithms. The other algorithms shown in the plot are: stochastic MTFM decoding algorithm, IDB algorithm, GDBF algorithm and the Offset Min-Sum algorithm. From the plot, it could be observed that the NGDBF algorithm performs better than the IDB algorithm. The IDB algorithm reaches an error rate of $10^{-7}$ at an $E_{b}/N_{0}$ of \SI{4.5}{\decibel}, while the NGDBF decoder is able to reach the same error rate at an $E_{b}/N_{0}$ of \SI{4.45}{\decibel}, similar to the stochastic decoder. In the case of the IDB decoder,  the failed frames are re-decoded six more times, with maximum number of iterations limited to $45$ for each phase. 

Fig. \ref{Fig:Avg_IEEE802.3an} shows the average number of iterations taken by the NGDBF decoder to converge with variation in $E_{b}/N_{0}$. From the plot, it could be seen that the IDB decoder has faster convergence compared to the NGDBF decoder. NGDBF converges faster compared to the stochastic MTFM decoder. With increase in $E_{b}/N_{0}$, the gap in the average number of iterations between NGDBF and IDB reduces.

\begin{figure*}[t!]
  \hspace*{0.285cm}     
  \includegraphics[scale = 1.1]{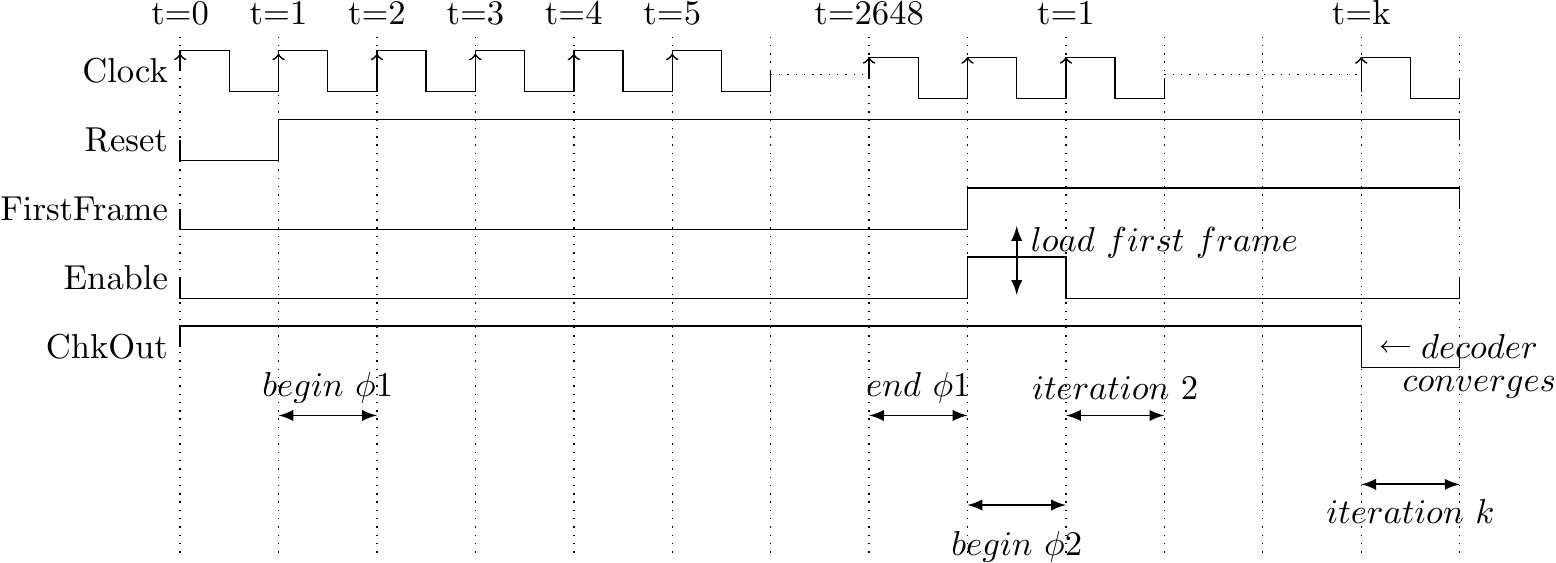}
  \caption{Timing diagram of the decoder. The start of the two operational phases is clearly shown. At $t=2649$, the decoding operation begins and the first frame is loaded into the decoder.}
  \label{Fig:Waveform}
\end{figure*}

\section{Architecture of NGDBF Decoder}
\label{sec:Arch_IEEE_802.3an}

Fig. \ref{Fig:Top_Arch} shows the top level architecture of the NGDBF decoder. The decoder has a fully parallel architecture and adopts a flooding schedule. The decoder consists of five main blocks: Noise Update Unit (NUU), Symbol Node Unit (SNU), Check Node Unit (CNU), Early Termination Unit (ETU) and the interleaver network. The check nodes are updated first and the symbol nodes are updated later during a decoding iteration. A decoding iteration takes a clock cycle. The fully parallel NGDBF decoder has 2048 symbol node processors and 384 check node processors. 
The decoder is operated in two phases: The first phase $\phi1$ is termed as the start-up phase. In this phase, noise samples are obtained, processed and are stored in a set of registers. The operation in this phase only involves the NUU. During the second phase $\phi2$, decoding operation is initiated and the decoder starts to decode. ETU detects convergence, signals the need for the current frame to be removed and a new frame to be loaded in. Fig. \ref{Fig:Waveform} shows the waveform diagram corresponding to the startup phase of the decoder. 
During the first $2648$ cycles, standard Gaussian samples of width $Q$ bits are loaded in from the input {\em Noisein} at the rate of one sample every clock cycle. The NUU processes each sample and stores them in a $Q-1$ bit register. After the end of $2648$ clock cycles, the first frame is loaded and decoding starts. Every iteration takes a clock cycle. Decoding throughput of an LDPC decoder can be calculated as follows:
\begin{equation}
\textrm{Throughput} ={\frac{f N}{st} } 
\end{equation}
where $f$ is the maximum speed of the decoder and is determined by the latency of one iterative check and symbol node processing. Since the decoder completes one iteration per cycle, $s$ = 1. Table \ref{tbl:Wavefromcontrol} shows the control logic for synchronizing and controlling all the phases of the decoder operation. At $t=k$, the {\em ChkOut} is low and the decoder converges. 

\subsection{NUU Design}
\label{sec:Arch_NUU}

We now consider the algorithm's implementation with quantized arithmetic. Retaining the notation from \cite{Sundararajan_2014_TCOMM}, we use $\tilde{y}$ to represent the quantized value of some signal $y$. Then the calculation performed at symbol node $k$ is given by

\begin{equation}
\tilde{E}_k\left( t \right) = x_k\left( t \right) \tilde{y}_k + \tilde{w}\sum_{i\in \mathcal{M}\left(k\right)} s_i + \tilde{q}_k\left( t \right).
\end{equation}
The symbol node evaluates the right-hand side of the above equation and then flips the decision bit ($x_{k}$), if $\tilde{E_{k}}$ is less than inversion threshold $\tilde{\theta}$.  This is done by calculating the sign of the below equation:
\begin{equation}
\tilde{E}_k\left( t \right) - \tilde{\theta} = x_k\left( t \right) \tilde{y}_k + \tilde{w}\sum_{i\in \mathcal{M}\left(k\right)} s_i + \tilde{q}_k\left( t \right) - \tilde{\theta}.
\label{eq:dec}
\end{equation}

\begin{table}[!t]
	\begin{center}
		\caption{Control table.}
		\label{tbl:Wavefromcontrol}		
		\begin{tabular}{|c|c|c|}
			\hline
			$\textbf{FirstFrame}$ & $\textbf{Enable}$ & $\textbf{Operation}$ \\
			\hline
			$0$ & $0$ & $\phi1$ begins \\ \hline
			$1$ & $1$ & $\phi2$ begins and new frame loaded \\  \hline
			$1$ & $0$ & Decoding iteration begins \\ 	\hline
		\end{tabular}
	\end{center}
\end{table}

\begin{figure*}[!t]
  \hspace*{1.7cm}     
  \includegraphics[scale = 0.9]{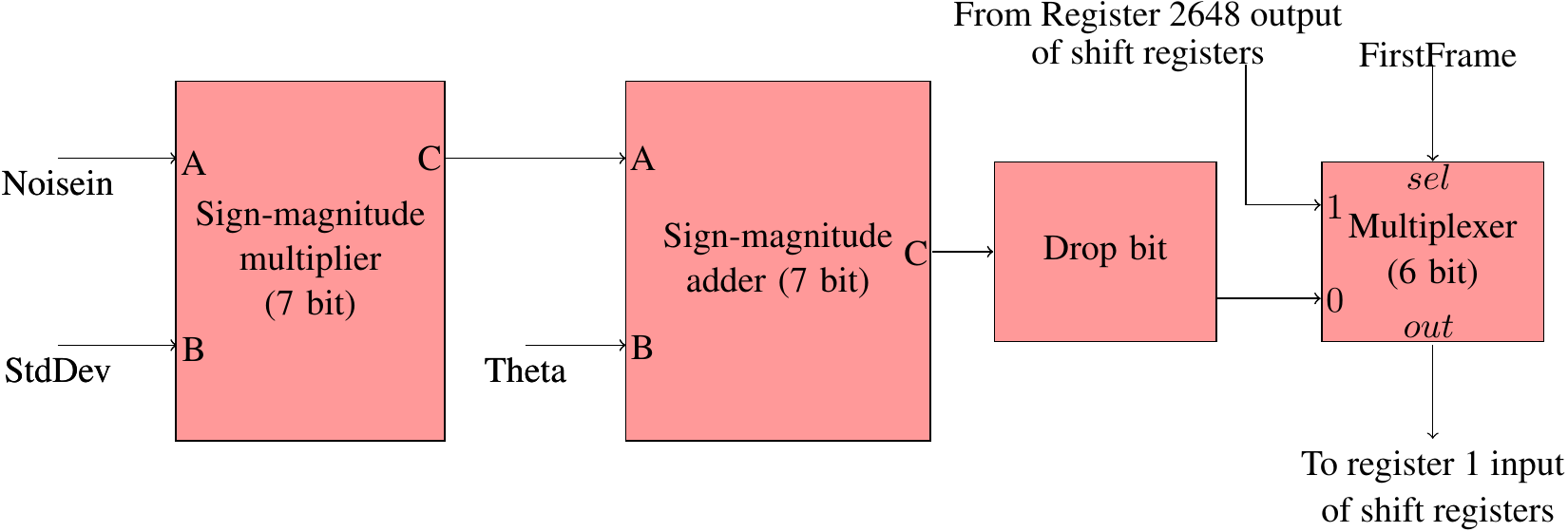}
  \caption{Architecture of NUU.}
  \label{Fig:NUU}
\end{figure*}

The right-hand side of equation (\ref{eq:dec}) contains four terms. Only the first two terms involve quantities that will be updated during a decoding iteration. The first term involves current hard decision and the second term involves summation of the syndromes that are obtained from the neighboring check nodes. The third term and the fourth term involve operations that are independent of either symbol node or check node updates and could be done prior to the start of decoding to reduce the decoding latency. In this design, these operations are done in the start-up phase by the NUU. Fig. \ref{Fig:NUU} shows the architecture of the NUU. As discussed in \cite{Sundararajan_2014_TCOMM}, noise generation can be simplified by generating all the samples during the start and then reusing them by just shifting the noise samples from one register to another during a decoding iteration. However, the architecture described in \cite{Sundararajan_2014_TCOMM} still requires one Gaussian random number generator. Gaussian random number generators are very complex and incur large complexity both in space and in time \cite{Lee_2005,Malik_2013,Danger_2000}. 

\begin{figure}[]
  \hspace*{1.7cm}    
	\begin{center}
          \includegraphics[scale = 0.7]{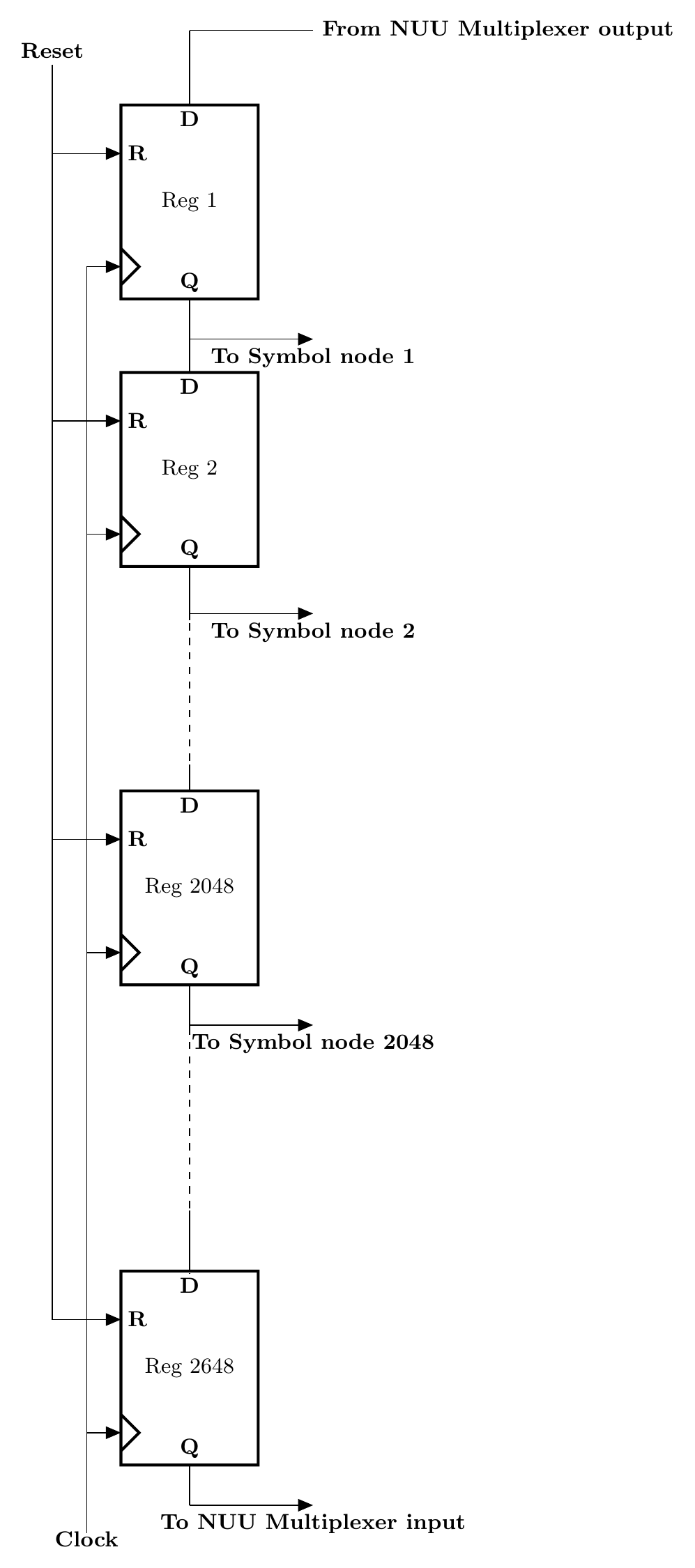}
          \caption{Shift registers containing noise samples.}
          \label{Fig:SR}
	\end{center}
\end{figure}

As an alternative to the architecture described in \cite{Sundararajan_2014_TCOMM}, a noise generation method is proposed in this section that is more efficient in comparison and does not require an on-chip Gaussian random number generator. The NUU consists of a seven bit sign-magnitude multiplier, a seven bit sign-magnitude adder and a set of $2648$ $Q-1$ bit shift registers. During the start-up phase, the NUU receives samples from the input {\em Noisein}. The input is a Gaussian distribution noise sample with zero-mean and unit variance. All the received samples are in sign-magnitude format. The total length of the sample is $Q=7$ bits, with one bit representing the sign, two bits representing the integer part and four bits representing the fractional part. The sample is then multiplied with the desired standard deviation obtained from {\em StdDev} input. The multiplier output is then added with the inversion threshold {\em Theta}. 

The Most Significant Bit (MSB) of the integer part of the resulting sum is dropped and the remaining six bits are written to the first 6 bit register Reg $1$ of a series of shift registers, as shown in Fig. \ref{Fig:SR}. The sixth bit in the sum corresponds to MSB of the integer part of the sum. This bit was mostly found to be zero from our simulations and is dropped without significantly affecting the error performance. Since the shift-register comprises a large portion of the design's area, dropping a bit provides a noticeable reduction in area and power dissipation. 

During the power-up phase, samples are scanned serially until all the $2648$ registers are loaded with noise samples that have the appropriate variance of $\eta \sigma^2$. During the phase $\phi2$, all the samples in the registers are circularly shifted from top to bottom. As shown in Fig. \ref{Fig:SR}, outputs of registers $1$-$2048$ are connected to the symbol nodes $1$-$2048$ respectively. Circular shifting allows for samples to be reused, and ensures that each symbol node locally receives a high-quality sequence of random samples. Although there is some correlation between different symbol nodes at different times, it has no apparent impact on performance.

\subsection{Symbol Node Design}
\label{sec:Arch_SNU}

\begin{figure*}[!t]
	\hspace*{1.1cm}      
	\begin{center}
	\includegraphics[scale = 0.85]{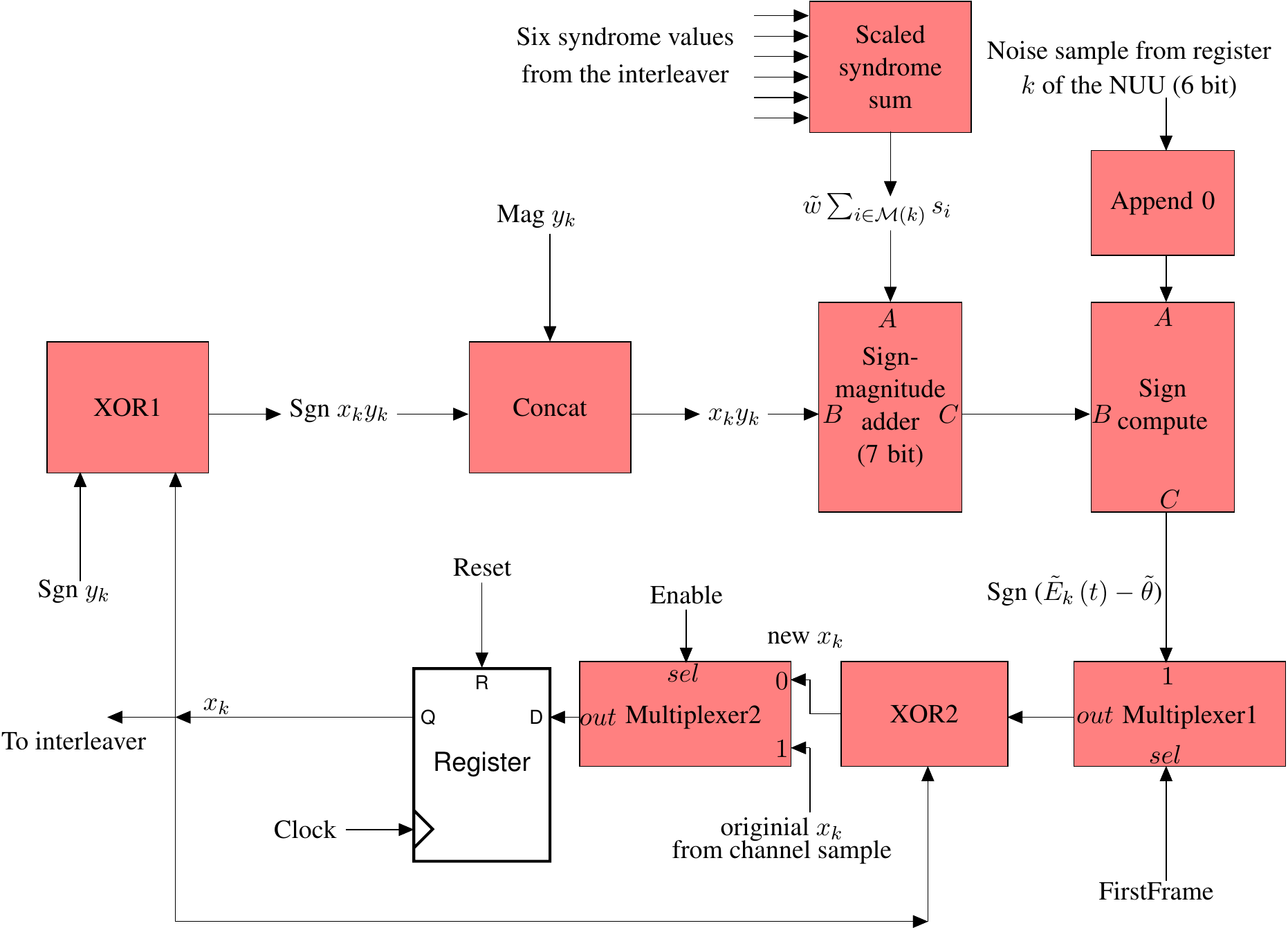}
        \caption{Architecture of a symbol node $k$. Mag corresponds to magnitude and Sgn corresponds to the sign.}
        \label{Fig:SNU}
      \end{center}
\end{figure*}

Fig. \ref{Fig:SNU} shows the architecture of the symbol node $k$. The node accepts a channel sample $y_{k}$ and computes the initial hard decision $x_{k}={\rm sign}\left(y_k\right)$. 
The decision is then passed on to the interleaver, which routes the hard decision to all the check nodes that are connected to the symbol node. 
The syndrome components are computed at the check nodes and are then transmitted to the symbol nodes via the interleaver routing. In the symbol node, the syndrome signals are passed on to the scaled syndrome sum module.

\begin{figure*}[!t]
  \begin{subfigure}{0.5\textwidth}
    \centering
    \includegraphics[scale=0.5]{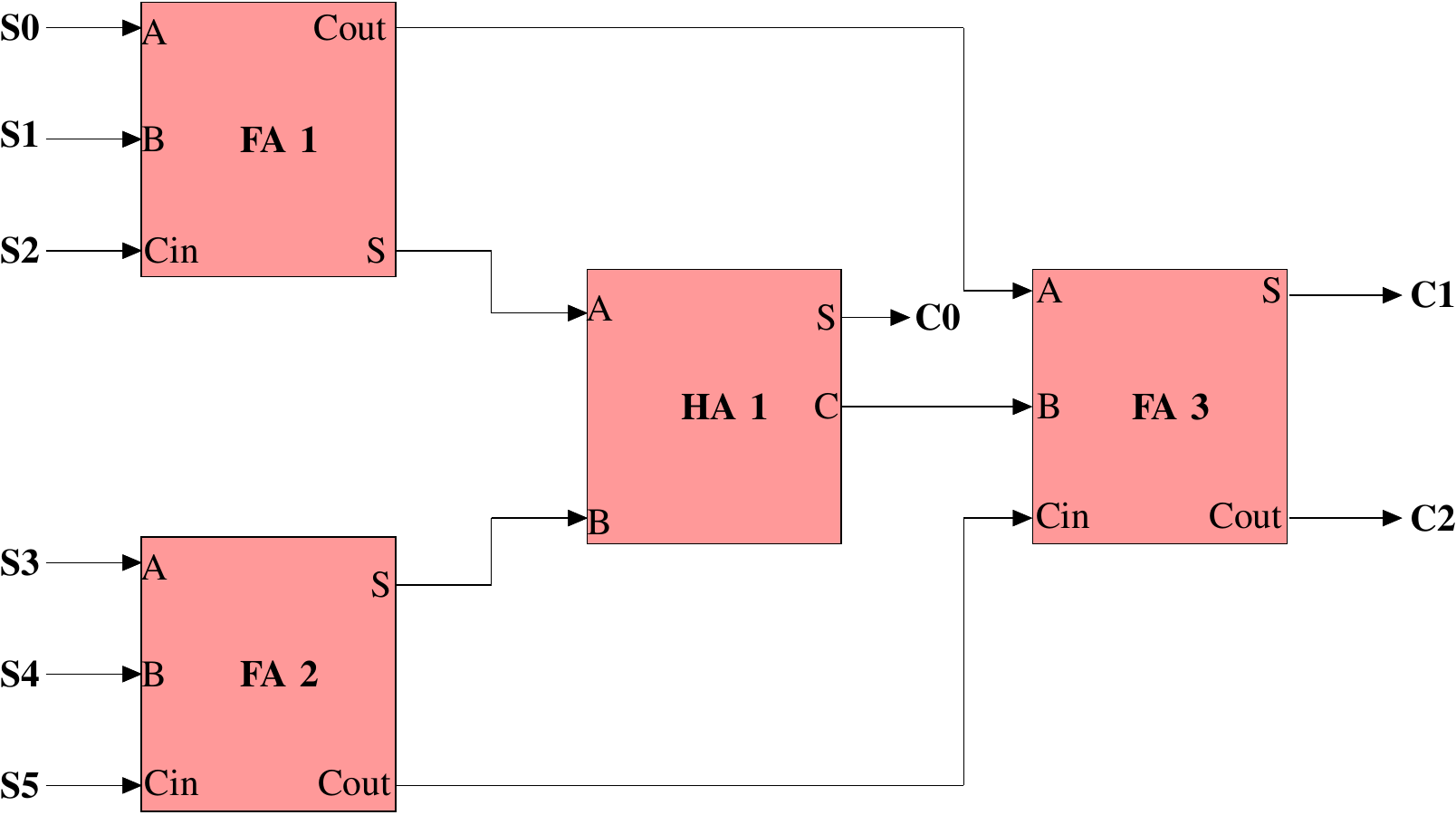}
    \caption{Architecture of count module. The module produces a three bit output ($C0-C2$). FA indicates a full adder and HA indicates a half adder.}		
    \label{Fig:lut}
  \end{subfigure}
  ~
  \begin{subfigure}{0.5\textwidth}
    \centering
    \includegraphics[scale=0.53]{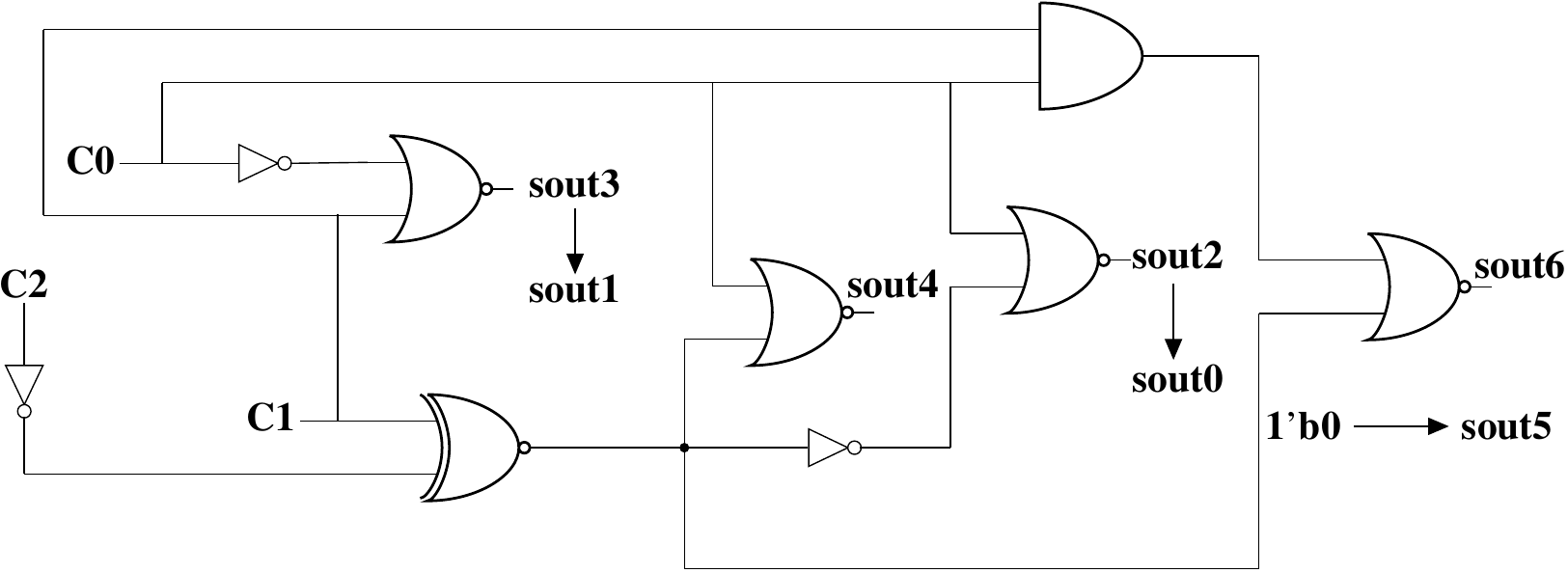}		
    \caption{Logic diagram of the lookup table.}
    \label{Fig:count}
  \end{subfigure}
  \caption{Scaled syndrome module. }
\end{figure*}

\begin{table}[!t]
	\begin{center}
		\caption{Scaled syndrome sum lookup table.}
		\label{tbl:Syndromesumscale}		
		\begin{tabular}{|c|c|c|}
			\hline
			$\textbf{Count}$ & $\textbf{Scaled syndrome sum values}$ & $\textbf{Sign-magnitude values}$ \\
			\hline
			$0$ & $6/6(1)$ & $0010000$ \\ \hline
			$1$ & $4/6(0.666)$ & $0001010$ \\ \hline
			$2$ & $2/6(0.333)$ & $0000101$ \\ \hline
			$3$ & $0/6(0)$ & $0000000$ \\ \hline	
			$4$ & $-2/6(-0.333)$ & $1000101$ \\ \hline
			$5$ & $-4/6(-0.666)$ & $1001010$ \\ \hline
			$6$ & $-6/6(-1)$ & $1010000$ \\ \hline											
		\end{tabular}
	\end{center}
\end{table}

The scaled syndrome sum module counts the number of ones in the incoming syndromes, computes the sum and scales down the syndrome sum by a factor of six. Fig. \ref{Fig:count} shows the architecture of the count part of the scaled syndrome module. The count module consists of three full adders and a half adder. The count module takes in all the six one bit syndromes and produces a three bit output that indicates the number of ones in the incoming syndromes. The number of ones can vary from zero to six. In the sign-magnitude numerical format, every count of one corresponds to a negative one and every count of zero corresponds to a positive one. Hence, the final output of the scaled syndrome sum can be obtained from Table \ref{tbl:Syndromesumscale}. Then according to the count, a look-up table is implemented that calculates the final output in the sign-magnitude format. The logic diagram of the look-up table is shown in Fig. \ref{Fig:lut}. The look-up table consists of three inverters, four NOR gates, one XNOR gate and one AND gate. The computed $x_{k}y_{k}$ and the scaled syndrome values are passed on as inputs to a seven bit sign-magnitude adder and the sum is passed on to the sign compute module where it is added with the noise sample. 
Since the objective is to calculate the final sign, the sign compute module does not need to calculate the final magnitude, which allows some reduction in complexity.
The output from the sign compute module indicates whether the bit should be flipped. The new decision is therefore obtained as the XOR of the previous decision with the sign output. This new decision that appears at the input of the register is made transparent on the next positive clock edge and is passed on to the interleaver and the XOR gates for the next iteration. All the above mentioned operations are repeated again.

\subsection{Check Node and ETU Design}
\label{sec:Arch_CNU}

The check node is an XOR gate that takes in 32 decisions from the neighboring symbol nodes and calculates the syndrome. The ETU detects convergence by computing the OR operation over all syndrome components. When all syndrome outputs are zero, the ETU asserts the {\em ChkOut} signal to indicate that decoding is complete.

\section{Implementation Results}
\label{sec:Results}

\begin{figure}[!t]
  \hspace*{1.4cm}      
  \begin{center}		
    \includegraphics[scale = 0.25]{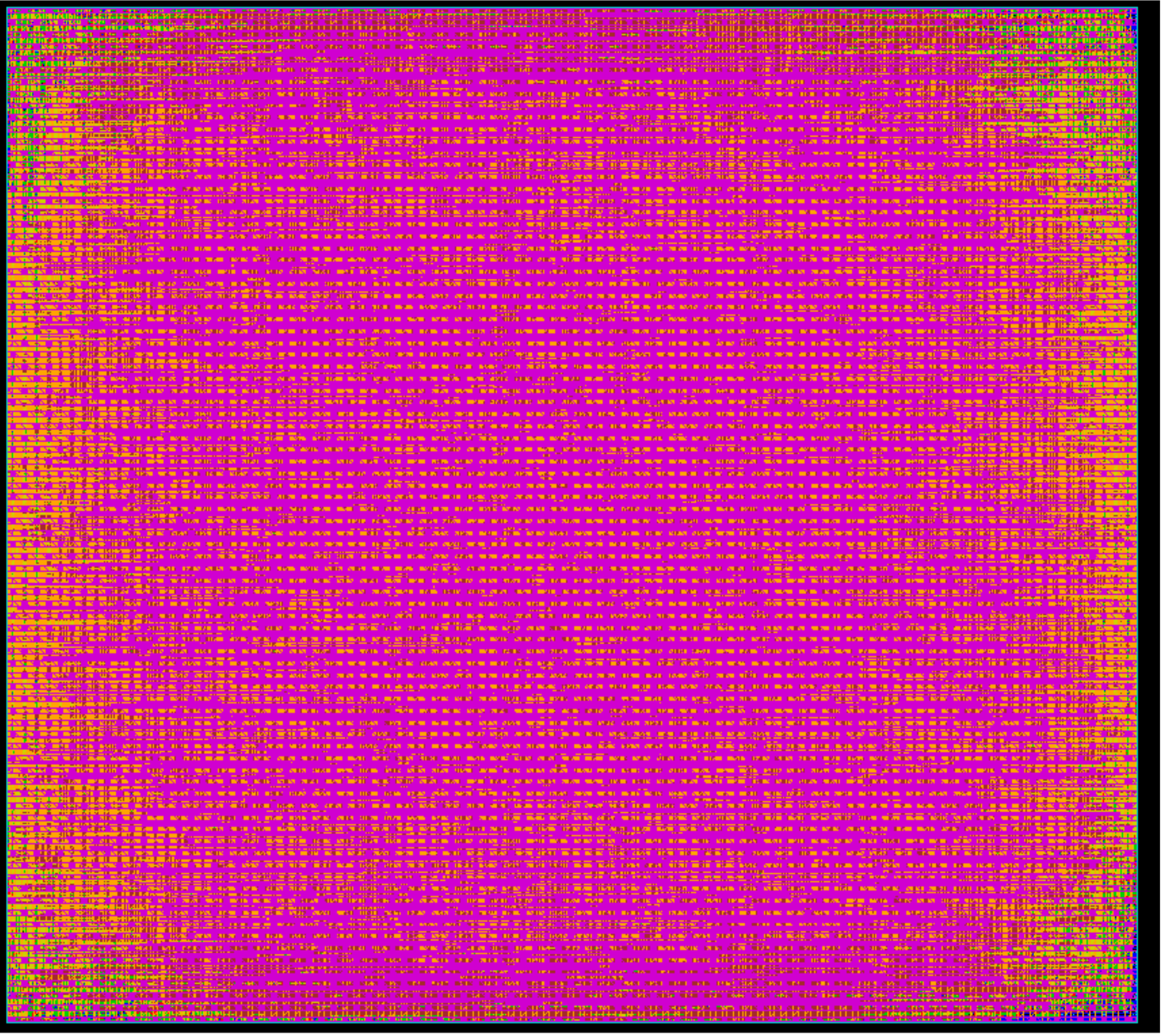}
    \caption{Layout view of NGDBF decoder implementation.}
    \label{Fig:Decoder_implement}
  \end{center}
\end{figure}

This section describes the implementation results and compares all the parameters of the NGDBF decoder to other existing state of the art decoders. The decoder design was implemented in Verilog and synthesized using Synopsys Design Compiler. Place and Route was performed using Cadence SOC Encounter. Synopsys Primetime was used for post-layout timing analysis and power analysis. The designs were synthesized using commercial \SI{65}{\nano\metre} standard cell libraries from ST Micro. All the presented results use nominal operating conditions. Nominal operating conditions correspond to nominal process corner, supply voltage of \SI{1}{\volt} and temperature of \SI{25}{\celsius}. Fig.~\ref{Fig:Decoder_implement} shows the layout of the final routed design. The total wire-length of NGDBF decoder is \SI{7.37}{\metre}, while the total wire-length of the IDB decoder is \SI{10.95}{\metre}, i.e.\ the wire-length of the NGDBF decoder is $0.672$ times smaller compared to the IDB decoder. 

\begin{figure}[!t]
	\begin{center}
          \includegraphics{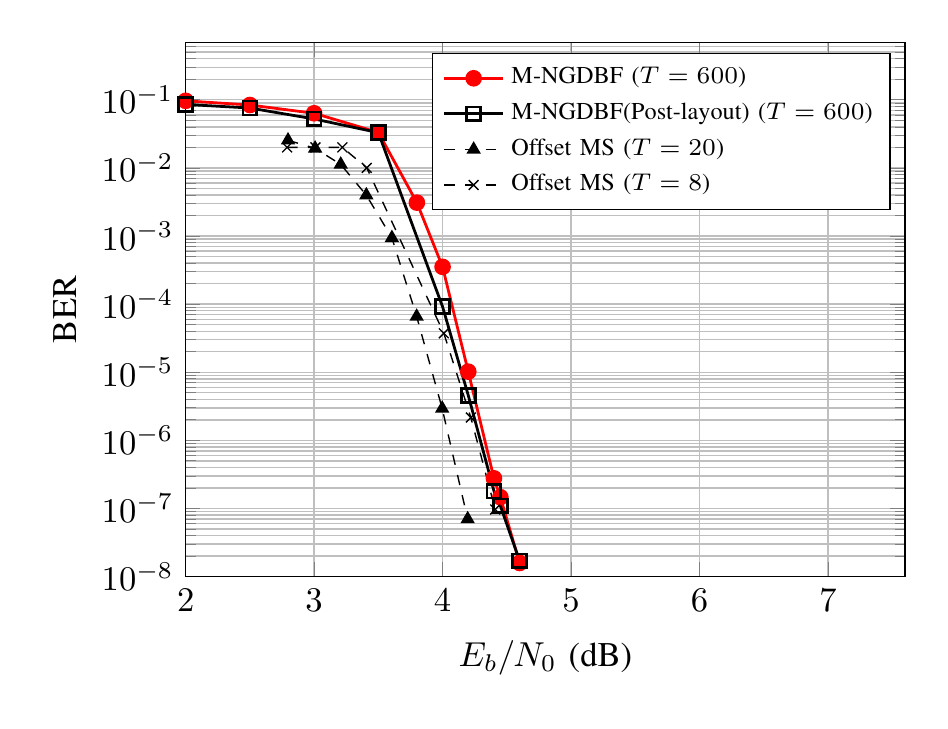}
		\caption{ BER for routed NGDBF compared to a benchmark OMS decoder for the IEEE 802.3 standard LDPC code with maximum iterations limited to $T$.}
		\label{Fig:BER_IEEE802.3an_PL}
	\end{center}
\end{figure}

\begin{figure}[]
  \begin{center}
    \includegraphics{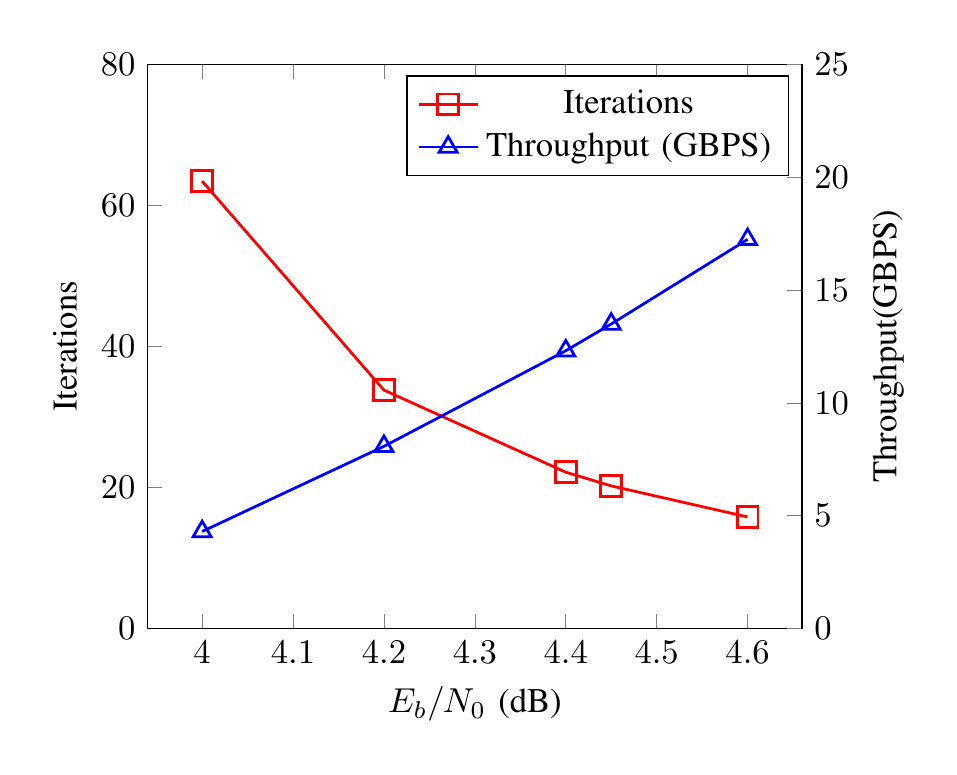}		
    \caption{Average number of iterations and average throughput versus $E_{b}/N_{0}$ for routed NGDBF design for IEEE 802.3an standard LDPC code with maximum iterations limited to $T$.}
    \label{Fig:Avg_IEEE802.3an_PL}
  \end{center}
\end{figure}

\begin{figure}[]
  \begin{center}
    \includegraphics{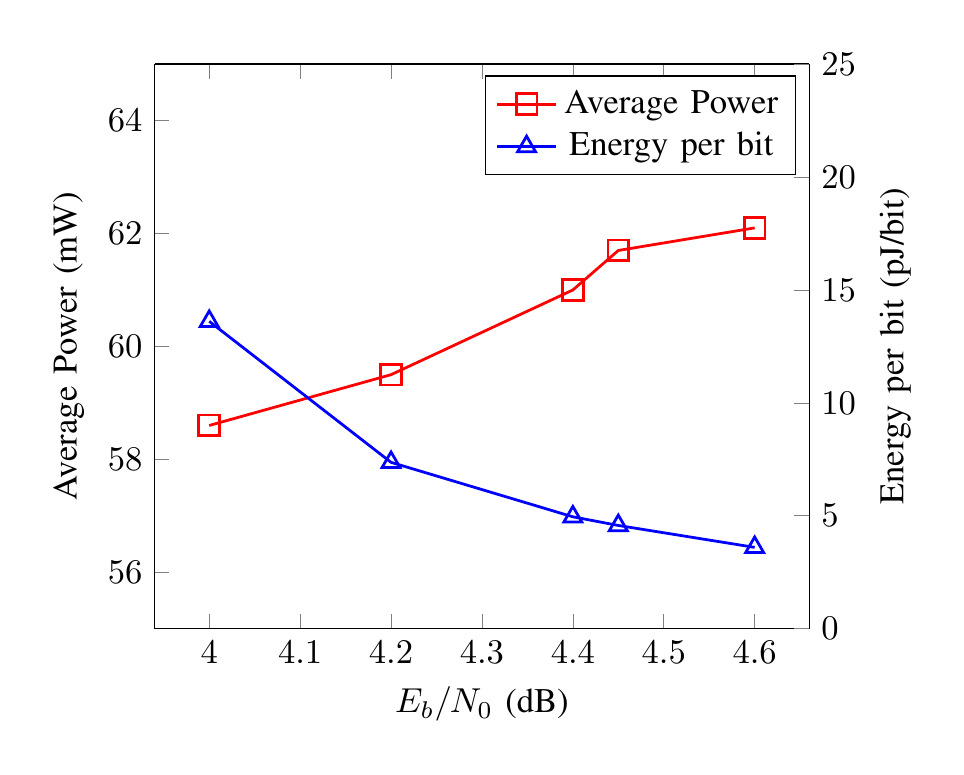}
    \caption{Energy per bit consumption for routed NGDBF design for IEEE 802.3 standard LDPC code with variation in $E_{b}/N_{0}$.}
    \label{Fig:Energy}
  \end{center}
\end{figure}

Fig. \ref{Fig:BER_IEEE802.3an_PL} shows the BER results obtained from the post-layout functional simulations. From the plot, it is observed that the post-layout simulation matches closely with the system level simulation. The NGDBF decoder is able to achieve an error rate of $10^{-7}$ at an $E_{b}/N_{0}$ of $4.45$ $\si{\deci\bel}$. Fig. \ref{Fig:Avg_IEEE802.3an_PL} shows the average number of iterations taken by the routed NGDBF decoder to converge. The average throughput obtained is also plotted. Throughput is low at lower values of $E_{b}/N_{0}$ and increases with increase in $E_{b}/N_{0}$. The maximum frequency at which the decoder could operate successfully was estimated to be \SI{133.33}{\mega\hertz} ($T=$\SI{7.5}{\nano\second}). At this frequency, the decoder crosses the required minimum throughput of \SI{10}{\giga b\per\second} at an $E_{b}/N_{0}$ of $4.3$ $\si{\deci\bel}$. At $E_{b}/N_{0}$ of $4.45$ $\si{\deci\bel}$, where the decoder attains an error rate of $10^{-7}$, the average throughput is \SI{13.5}{\giga bp\second}. Fig. \ref{Fig:Energy} shows the plot of energy per bit consumed and the average power dissipated at different values of $E_{b}/N_{0}$. Energy per bit is obtained as the ratio of average power to average throughput. Energy per bit decreases with increase in $E_{b}/N_{0}$. This is due to increase in throughput with reduction in average number of iterations. 

\begin{figure*}[!t]
	\hspace*{1.2cm}      
	\includegraphics[scale = 0.85]{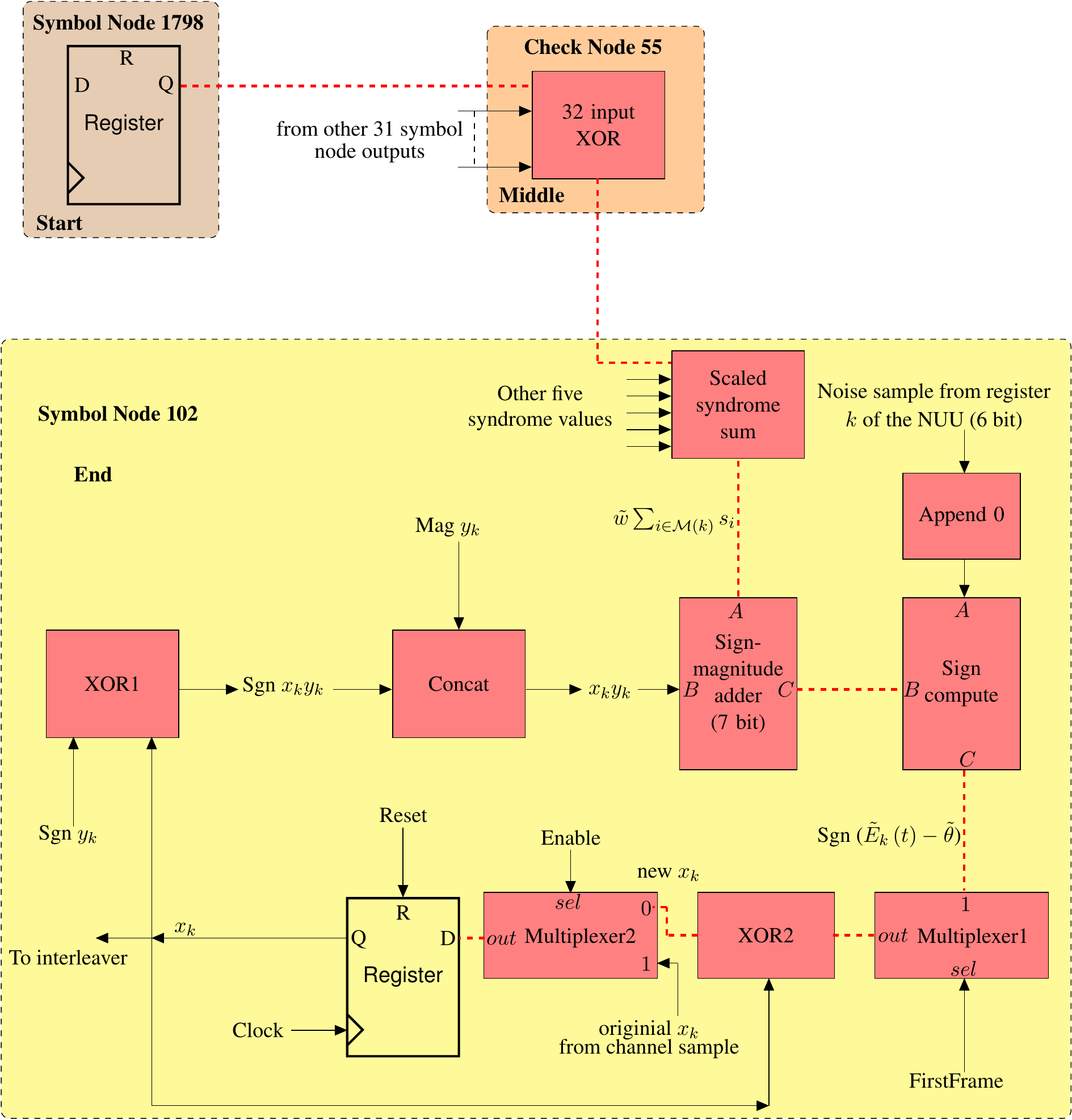}
	\begin{center}
		\caption{Critical path showing the start, middle and the end blocks. The actual critical path is highligted using thick broken lines.}
		\label{Fig:CP}
	\end{center}
\end{figure*}

\begin{table}[]
	\begin{center}
		\caption{NGDBF critical path.}
		\label{tbl:cpath}		
		\begin{tabular}{|c|c|}
			\hline
			$\textbf{Component}$ & $\textbf{Delay(ns)}$  \\
			\hline 	\hline
			Symbol Node 1798 clock buffers & $0.564$ \\ \hline
			Symbol Node 1798 clock-Q 	   & $0.254$ \\ \hline
			Symbol Node 1798 to Check Node 55 buffers 	   & $0.287$ \\ \hline		
			Check Node 55 Xor gates	   	   & $2.12$ \\ \hline	
			Check Node 55 buffers to Symbol Node 102 buffers & $0.69$ \\ \hline	
			Symbol Node 102 combinational logic & $3.58$ \\ \hline	\hline	
			\textbf{Total Delay}& $\textbf{7.495}$ \\ \hline												
		\end{tabular}
	\end{center}
\end{table}

Fig. \ref{Fig:CP} shows the critical path of the routed design. In any synchronous design, the critical path always starts at the output of a register, propagates through a set of combinational logic and ends at the input of a register. In the case of our decoder, the critical path starts at the output of the register in the start symbol node and ends at the input of the register at the destination symbol node. Based on that, the critical path of the decoder can be split into three main sub-paths. The first sub-path involves the decision register of the start symbol node, buffers between the start symbol node and the adjacent check node. The second sub-path involves the check node and the buffers between the check node and the destination symbol node. The third sub-path involves the combinational logic in the destination symbol node that ends at the input of the destination register. The start symbol node is symbol node $1798$. The check node is check node $55$. The terminating symbol node is symbol node $102$. 
Table \ref{tbl:cpath} shows the contribution of each component to the critical path delay. 
The adders represent a significant contribution to the symbol node delay. The total delay of adders present  in the syndrome scale unit and in the sign-magnitude adder unit is \SI{1.068}{\nano\second}, which represents about 30\% of the symbol node's total delay. The critical path is highlighted in Fig.~\ref{Fig:CP} by thick broken lines. 

The implementation results for the NGDBF decoder are summarized in Table \ref{tbl:Imp}, along with other works on the 10GBASE-T Ethernet decoder. The NGDBF decoder attains significant improvements in area, error rate and energy efficiency. The area of the routed NGDBF design is \SI{0.81}{\milli\metre^2}. Among the previous works listed in Table \ref{tbl:Imp}, the IDB decoder has the smallest area at \SI{1.44}{\milli\metre^2}. The NGDBF decoder occupies $43.7$\% smaller silicon area than the IDB decoder. It also occupies $75$\%  smaller area than stochastic MTFM decoder, which has the second smallest area. Hence, the proposed NGDBF decoder outperforms other existing state of the art decoders in terms of area.

At $E_{b}/N_{0}$ = $4.55$ $\si{\deci\bel}$, the average throughput of the NGDBF decoder is \SI{14.6}{\giga bp\second}. This is lower than the IDB and the split-row MS decoders. This is because, at this low SNR, NGDBF requires more iterations on average than the IDB and the split-row MS algorithms. The NGDBF decoder nevertheless exceeds the standard's requirement of \SI{10}{\giga bp\second}, and has a lower average power consumption of \SI{61.6}{\milli\watt}. This is much lower than the IDB and the split-row MS decoders. To obtain a normalized comparison with the benchmark decoders, we consider the energy per bit and throughput per unit area, which are useful figures of merit for comparing decoder architectures \cite{Kienle_TCOMM_2011}. At $E_{b}/N_{0}$ = $4.55$ $\si{\deci\bel}$, the NGDBF decoder is second best after IDB in terms of energy efficiency, and is $2.05$ times better than the normalized energy per bit of the split-row MS design. In terms of throughput per unit area, both the IDB and split-row MS outperform the NGDBF decoder. The throughput per unit area of the NGDBF decoder is $4.3$ times better than layered offset MS decoder at $E_{b}/N_{0}$ = $4.55$ $\si{\deci\bel}$.

At a somewhat higher SNR of $E_{b}/N_{0}$ = $5.5$ $\si{\deci\bel}$, the NGDBF decoder becomes more favorable compared to the other designs. The average throughput of the NGDBF decoder is \SI{36.4}{\giga bp\second}. This is lower than the IDB, stochastic MTFM and the offset MS decoders. However, the NGDBF decoder's average power consumption increases only slightly, to \SI{63}{\milli\watt}. At this SNR, the NGDBF becomes the most energy efficient decoder, having an energy per bit that is $1.6$ times better than IDB design and $23.52$ times better than the normalized energy per bit of the offset MS design. This energy efficiency is the lowest reported in the research literature for 10GBASE-T Ethernet decoders. NGDBF has the second best throughput per unit area after IDB and outperforms the offset MS, layered offset MS decoder and the stochastic decoder at this SNR. The Offset MS decoder represents a very common architecture for commercial LDPC decoders, and the NGDBF decoder has $10.73$ times better throughput per area than the Offset MS benchmark.

In terms of BER performance, the offset MS decoder has the best coding gain, but is comparatively very complex and incurs a large area overheard. Among the  other simplified implementations, NGDBF has the best performance. NGDBF achieves a gain of \SI{0.05}{\deci\bel} compared to the IDB and \SI{0.1}{\deci\bel} compared to split-row MS. Even though the stochastic MTFM algorithm has a much higher complexity compared to the NGDBF algorithm, NGDBF takes the same $E_{b}/N_{0}$ as the stochastic MTFM to reach a BER of $10^{-7}$. 

\begin{table*}[]
	\renewcommand\arraystretch{0.76}
		\caption{Implementation results for NGDBF and comparison with other works.}
		\label{tbl:Imp}
		
		\hspace{0.3cm}
	    \begin{minipage}{\textwidth}	
	    	
	    	\scalebox{1.17}{	
		\begin{tabular}{|c|c|c|c|c|c|c|}
		   \hline
		   $\textbf{Parameter}$ & $\textbf{This Work}$ &  \cite{Cushon_2014} &  \cite{Mohsenin_2010a} &  \cite{Tehrani_2010} &  \cite{Zhang_2010a} &  \cite{Cevrero_2010}  \\
		   \hline \hline
           \begin{tabular}{@{}c@{}}\textbf{Decoding} \\ \textbf{algorithm}\end{tabular} & M-NGDBF & IDB & \begin{tabular}{@{}c@{}}Split-Row  \\ MS\end{tabular} & \begin{tabular}{@{}c@{}}Stochastic \\ MTFM\end{tabular} & \begin{tabular}{@{}c@{}}Offset\\ MS\end{tabular} & \begin{tabular}{@{}c@{}}Layered\\ Offset MS\end{tabular}  \\ \hline
           \textbf{Technology} & $65$ $nm$ & $65$ $nm$ & $65$ $nm$  & $90$ $nm$  & $65$ $nm$  & $90$ $nm$   \\  \hline 
           \begin{tabular}{@{}c@{}}\textbf{Quantization} \\ \textbf{bits}\end{tabular} & $7$ & $6$  & $5$   & $6$   & $4$  & $4$   \\         \hline   
           \begin{tabular}{@{}c@{}}\textbf{Area (scaled} \\ \textbf{to $\textbf{65}$ $\textbf{nm}$) ($\textbf{mm}^\textbf{2}$)}\end{tabular} & $0.81$ & $1.44$  & $4.84$   & $6.38$ $(3.33)$   & $5.35$  & $5.35(2.79)$     \\         \hline          
           \begin{tabular}{@{}c@{}}\textbf{Maximum} \\ \textbf{iterations}\end{tabular} & $600$ & $315$  & $11$ & $400$  & \begin{tabular}{@{}c@{}}$8 + 6$ \\ post-processing\end{tabular} & $4$ \\ \hline  
           \begin{tabular}{@{}c@{}}\textbf{Area} \\ \textbf{Utilisation(\%)}\end{tabular} & $92.2$ & $95$ & $97$ & $95$ & $84.5$ & $84.4$ \\ \hline               
            \begin{tabular}{@{}c@{}}\textbf{$\textbf{E}_{\textbf{b}}/\textbf{N}_{\textbf{0}}$ at} \\ \textbf{BER = $\textbf{10}^{\textbf{-7}}$ }\end{tabular} & $4.45$ & $4.5$ & $4.55$ & $4.45$ & $4.25$ & $4.4$    \\  \hline        
           \begin{tabular}{@{}c@{}}\textbf{Supply} \\ \textbf{voltage ($\textbf{V}$)}\end{tabular} & $1.0$ & $1.0$  & $1.3$   & $1.0$   & $1.2$ &  $1.2$   \\         \hline                                                 
           \begin{tabular}{@{}c@{}}\textbf{Clock} \\ \textbf{frequency ($\textbf{Mhz}$)}\end{tabular} & $133.33$ & $520$  & $195$   & $500$   & $700$ &  $137$    \\         \hline  	
           \begin{tabular}{@{}c@{}}\textbf{Minimum} \\ \textbf{throughput ($\textbf{Gbps}$)}\end{tabular} & $0.46$ & $3.38$  & $36.3$   & $2.56$   & $14.9$  &  $11.7$    \\         \hline \hline   	
           
           \textbf{At $\textbf{E}_{\textbf{b}}/\textbf{N}_{\textbf{0}}$ = $\textbf{4.55}$ $\textbf{dB}$ }    \\  \hline 
           
           \begin{tabular}{@{}c@{}}\textbf{Average power} \\ \textbf{($\textbf{mW})$ }\end{tabular} & $61.6$ & $462$  & $1359$   & $-$   & $-$  &  $-$  \\         \hline  
           
           \begin{tabular}{@{}c@{}}\textbf{Average throughput} \\ \textbf{($\textbf{Gbps})$}\end{tabular} & $14.6$ & $126.3$  & $92.8$   & $-$   & $-$  & $11.7$   \\         \hline     
           
		   \begin{tabular}{@{}c@{}}\textbf{Energy per } \\ \textbf{bit ($\textbf{pJ/bit})$}\end{tabular} & $4.21$ & $3.65$  & $14.6$   & $-$   & $-$  &  $-$   \\         \hline   
		   
		   \begin{tabular}{@{}c@{}}\textbf{Throughput per scaled area  } \\ \textbf{($\textbf{Gbps/mm}^2)$}\end{tabular} & $18.02$ & $87.7$  & $19.2$   & $-$   & $-$ &  $2.18$    \\         \hline   \hline 		   
		   
           \textbf{At $\textbf{E}_{\textbf{b}}/\textbf{N}_{\textbf{0}}$ = $\textbf{5.5}$ $\textbf{dB}$ }    \\  \hline 
		   
		   \begin{tabular}{@{}c@{}}\textbf{Average power} \\ \textbf{($\textbf{mW})$}\end{tabular} & $63$ & $478$  & $-$   & $-$   & $2800$ & $-$     \\         \hline  		     
		   
		   \begin{tabular}{@{}c@{}}\textbf{Average throughput} \\ \textbf{($\textbf{Gbps})$}\end{tabular} & $36.4$ & $171.8$  & $-$   & $61.3$   & $47.7$  &  $11.7$  \\         \hline   
		   
		   \begin{tabular}{@{}c@{}}\textbf{Energy per } \\ \textbf{bit ($\textbf{pJ/bit})$}\end{tabular} & $1.73$ & $2.78$  & $-$   & $-$   & $58.7$ & $-$   \\         \hline    
		   
		   \begin{tabular}{@{}c@{}}\textbf{Throughput per scaled area  } \\ \textbf{($\textbf{Gbps/mm}^2)$}\end{tabular} & $44.94$ & $119.3$  & $-$   & $9.61$   & $8.92$  & $2.18$    \\         \hline   \hline 				   
		   		   	   
					   \begin{tabular}{@{}c@{}}\textbf{Scaled energy per bit } \\ \textbf{($\textbf{pJ/bit})$ at $\textbf{4.55}$ $\textbf{dB}$}\end{tabular} & $4.21$ & $3.65$  & $8.64$ \footnote{Normalized to $1.0$$V$} & $-$ & $-$ & $-$    \\         \hline  		   
					   
					   \begin{tabular}{@{}c@{}}\textbf{Scaled energy per } \\ \textbf{bit ($\textbf{pJ/bit})$ at $\textbf{5.5}$ $\textbf{dB}$}\end{tabular} \ & $1.73$ & $2.78$  & $-$   & $-$   & $40.8$ \footnote{Normalized to $1.0$$V$}  & $-$   \\         \hline  	   	\hline
					   
					   \begin{tabular}{@{}c@{}}\textbf{Scaled throughput} \\ \textbf{($\textbf{Gbps/mm}^\textbf{2})$ at $\textbf{4.55}$ $\textbf{dB}$}\end{tabular} & $18.02$ & $87.7$  & $19.2$ & $-$   & $-$ & $4.19$ \footnote{Area scaled to $65$ $nm$}    \\         \hline  		
					   		   
			   		   \begin{tabular}{@{}c@{}}\textbf{Scaled throughput} \\ \textbf{($\textbf{Gbps/mm}^\textbf{2})$ at $\textbf{5.5}$ $\textbf{dB}$}\end{tabular} & $44.94$ & $119.3$  & $-$   & $18.4$ \footnote{Area scaled to $65$ $nm$} & $8.92$  & $4.19$ \footnote{Area scaled to $65$ $nm$}   \\         \hline            	
		\end{tabular}	
	}
	    \end{minipage}	
\end{table*}

\begin{figure}[!t]
  \includegraphics{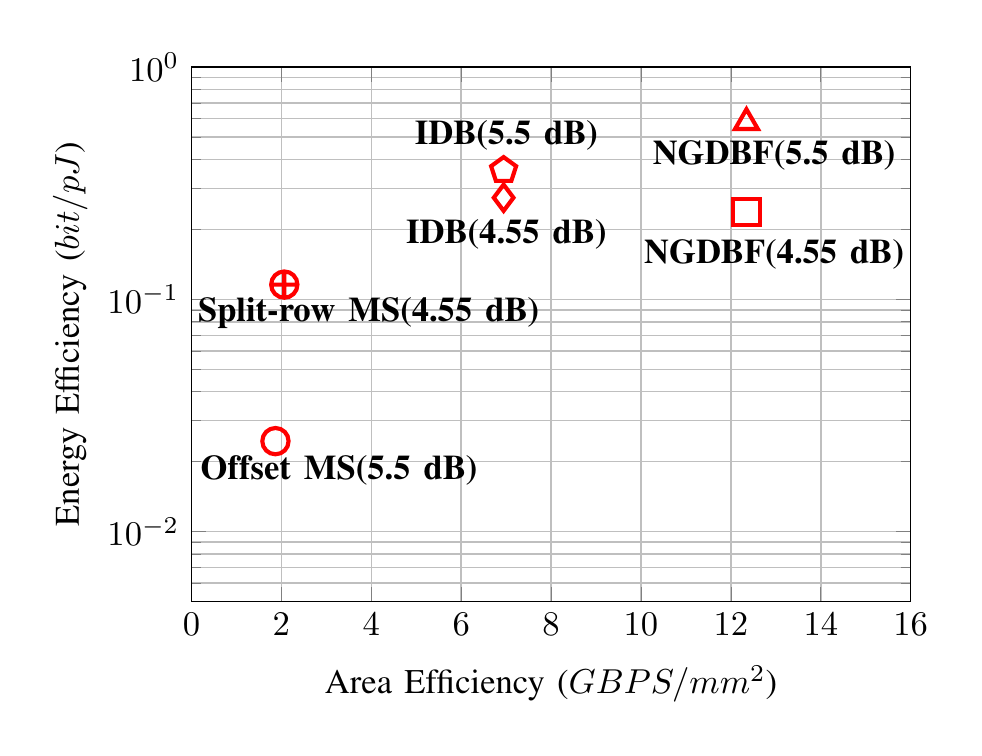}
  \caption{Energy efficiency versus area efficiency of all the reported $10$GBASE-T decoders. The energy values of Offset MS and Split-Row decoders are normalised to a supply voltage of $1.0$ V.}
  \label{Fig:FOM}
\end{figure}

Fig.~\ref{Fig:FOM} shows the plot of energy efficiency versus the area efficiency of all the reported 10GBASE-T Ethernet decoders operating at a minimum required throughput of $10$ \si{\giga bit\per\second}. We note that this scaling procedure is optimistic for the benchmark designs, since their efficiency would be diminished by leakage losses which are not accounted for in this normalization. In Fig.~\ref{Fig:FOM}, the less efficient designs appear toward the lower left corner, and the most efficient designs should appear near the upper right corner. From the plot, it can be seen that the NGDBF decoder operating at $E_{b}/N_{0}$ = \SI{5.5}{\deci\bel} is the most efficient decoder overall. The NGDBF decoder at  $E_{b}/N_{0}$ = $4.55$ $\si{\deci\bel}$ is second most efficient decoder. The offset MS decoder at $E_{b}/N_{0}$ = $5.5$ $\si{\deci\bel}$ is the least efficient decoder among these comparisons. The NGDBF decoder is expected to show continuing efficiency gains when operating at higher SNRs, since the average iterations per frame will decrease.

The main drawback of the NGDBF decoder is that it provides lower throughput compared to other state of the art decoders, and in particular there will be a small percentage of ``worst case'' frames which consume a large number of iterations, temporarily slowing the throughput to a level below \SI{10}{\giga bp\second}. But since NGDBF consumes a very low area, two or three instances of the decoder could be deployed for decoding, thereby increasing the throughput proportionally, and the total area would still be less than the standard Offset Min-Sum decoders. With multiple cores, it is also possible to achieve a second benefit, improving the BER performance by re-decoding, which has been shown to achieve significantly better coding gain than the Offset Min-Sum decoder \cite{journals/corr/TithiWS15}. Re-decoding is a phenomenon in which the failed frames are re-decoded in successive attempts with different noise values, thereby increasing the probability of decoding success. Even though re-decoding process consumes more energy, the energy dissipated should still be much less compared to the energy dissipated by the offset MS decoder. 

\section{Conclusions}
\label{Conclude}

An ASIC implementation of the NGDBF algorithm was implemented on a code that is deployed in 10GBASE-T Ethernet standard and the design was shown to be highly efficient both in terms of area and energy consumption at higher values of $E_{b}/N_{0}$ and was able to meet the standard's throughput requirements. Compared to previously reported implementations of the 10GBASE-T Ethernet decoder, the NGDBF decoder shows the best overall efficiency in terms of energy and area. Furthermore, compared to previous low-complexity implementations, the NGDBF decoder achieves better coding gain. 

\bibliographystyle{IEEEtran}
\bibliography{NGDBF_10GBaseT}

\end{document}